\documentclass[useAMS,usenatbib]{mn2e}

\usepackage[dvips]{graphicx}
\usepackage{times}
\usepackage{color}
\newcommand{\kms}{\mbox{$\mathrm{km\,s^{-1}}$}}
\newcommand{\mum}{\mbox{$\mathrm{\mu m}$}}
\newcommand{\Msun}{\mbox{${\rm M}_\odot$}}

\newcommand{\Rsun}{\mbox{${\rm R}_\odot$}}
\newcommand{\gtapprox}{\raisebox{-0.5ex}{$\,\stackrel{>}{\scriptstyle\sim}\,$}}
\newcommand{\ltapprox}{\raisebox{-0.5ex}{$\,\stackrel{<}{\scriptstyle\sim}\,$}}
\title[Balmer decrements in SS\,433]{SS\,433's circumbinary ring and
  accretion disc viewed through its attenuating disc wind}

\author[S. Perez and K. M. Blundell]{Sebastian Perez\thanks{E-mail:
    s.perez2@physics.ox.ac.uk } and Katherine M. Blundell\\University
  of Oxford, Department of Physics, Keble Road, Oxford, OX1 3RH, U.K.}

\begin{document}
\date{} 

\pagerange{\pageref{firstpage}--\pageref{lastpage}} \pubyear{2009}

\maketitle

\label{firstpage}

\begin{abstract}
  We present optical spectroscopy of the microquasar SS\,433 covering
  a significant fraction of a precessional cycle of its jet axis. The
  components of the prominent stationary H$\alpha$ and H$\beta$ lines
  are mainly identified as arising from three emitting regions: (i) a
  super-Eddington accretion disc wind, in the form of a broad
  component accounting for most of the mass loss from the system, (ii)
  a circumbinary disc of material that we presume is being excreted
  through the binary's L2 point, and (iii) the accretion disc itself
  as two remarkably persistent components. The accretion disc
  components move with a Keplerian velocity of $\gtapprox 600$~\kms\
  in the outer region of the disc. A direct result of this
  decomposition is the determination of the accretion disc size, whose
  outer radius attains $\sim$8~\Rsun\ in the case of Keplerian orbits
  around a black hole mass of 10~\Msun. We determine an upper limit
  for the accretion disc inner to outer radius ratio in SS\,433,
  $R_{\rm in}/R_{\rm out} \sim 0.2$, independent of the mass of the
  compact object. The Balmer decrements, H$\alpha/$H$\beta$, are
  extracted from the appropriate stationary emission lines for each
  component of the system. The physical parameters of the gaseous
  components are derived. The circumbinary ring decrement seems to be
  quite constant throughout precessional phase, implying a constant
  electron density of $\log N_{\rm e}({\rm cm}^{-3})\simeq 11.5$ for
  the circumbinary disc. The accretion disc wind shows a larger change
  in its decrements exhibiting a clear dependence on precessional
  phase, implying a sinusoid variation in its electron density $\log
  N_{\rm e}({\rm cm}^{-3})$ along our line-of-sight between 10 and
  13. This dependence of density on direction suggests that the
  accretion disc wind is polloidal in nature.
\end{abstract}

\begin{keywords}
  stars: individual: SS\,433 -- stars: winds, outflows -- binaries:
  spectroscopic
\end{keywords}

\section{Introduction}\label{sec:intro}

Microquasars are X-ray binaries which undergo a wide range of physical
processes including accretion onto a compact object (black hole or
neutron star) and the launch of relativistic jets. SS\,433, one of the
most studied microquasars, became famous as the first known source of
relativistic jets in the Galaxy, and it is the only system, X-ray
binary or active galactic nucleus (AGN), for which atomic emission
lines have so far been associated with the jets, hence implying a
baryonic content \citep[i.e., $e^- p^+$, ][]{mil79,cra81,fen00}.

The optical spectrum of SS\,433 is characterised by the presence of
numerous broad emission lines with complex profiles, on top of a
bright continuum. As at near-infrared and X-ray wavelengths, these
emission lines can be divided into two groups: lines that are referred
to as \textit{stationary}, albeit highly variable in strength and
profile, and lines that are \textit{moving}. The latter are thought to
originate in the two oppositely-directed relativistic jets moving with
a speed $v \sim 0.26~c$. The system SS\,433 shows four main
periodicities: the binary's orbital motion, with a period of about
13.08~days \citep{cra81}; the jet axis precession and nutation, with
periods of about 162 and 6~days \citet{kat82}, respectively, and a
recently discovered 550~day precession of the radio ruff
\citep{doo09}. A configuration where two jets emerge in opposite
directions from the central object \citep{fab79}, in which the jet
axis undergoes a precession cycle every 162~days was proposed to
describe the motion of the emission lines \citep{mil79}, and this is
referred to as the \textit{kinematical model}.

The stationary optical and near-infrared spectrum of SS\,433 is
dominated by hydrogen and {He\,\sc i} emission lines \citep{mar84} and
at least 15 per-cent of the flux in the hydrogen lines is contributed
by the accretion disc itself \citep{seba09}. The emission-line spectra
of accretion discs, and their evolution with orbital and precessional
phases, comprise most of the information we can obtain about the
temperature and density variations as well as velocity gradients
within the disc \citep{ski00}.

The Balmer decrements of the stationary lines in SS\,433 have not
previously been studied due to the high interstellar extinction
towards the object, which makes the detection of the H$\beta$ line
rather difficult \citep{pan97}. The decrements are highly dependent on
physical parameters of the gas such as its temperature, optical depth
and also the nature of the source of radiation \citep{dra80}.

The hard radiation field around compact objects and the expected high
electron densities for SS\,433 \citep[$N_{\rm e} \ge
10^{13}$~cm$^{-3}$, ][]{pan97} require that we use an adequate
treatment for such environments in order to study the emission-line
gas. At high densities, and in the presence of heavy elements,
excitation by collisional processes become a relevant factor
\citep{fer88}. \citet{dra80} performed theoretical calculations of the
emission-line spectrum from a slab of hydrogen at moderate to high
densities ($10^8 < N_{\rm e} < 10^{15}~$cm$^{-3}$) over a wide range
of physical parameters, including values close to those observed in
SS\,433's gas \citep{pan97}. We compare our estimates with Drake \&
Ulrich's findings in Section~\ref{sec:results}.

The existence of dust mixed with the emitting gas would have effects
on the Balmer decrement that are by no means negligible
\citep{ost06}. SS\,433's optical spectrum reveals evidence of severe
dust extinction, such as the remarkably red continuum and the presence
of prominent interstellar absorption lines in the form of diffuse
interstellar bands \citep{mar84}. In our optical data described in
subsequent sections, and in agreement with previous observations
\citep[e.g.,][among others]{mur80,mar84,gies02}, we have detected the
diffuse interstellar bands at 4430, 5778 and 5780~\AA\ and also the
interstellar lines Ca~H\,$\lambda3968$, Ca~K\,$\lambda3934$ as well as
Na~D\,$\lambda5890$. \citet{mur80} reported an interstellar absorption
(i.e., Galactic extinction) of $A_V \sim 8$~mag toward SS\,433,
obtained from infrared measurements. This result has been corroborated
by \citet{gies02} by fitting the spectral energy distribution, and the
currently most accepted value is $A_V = 7.8$~mag.

\begin{figure}
  \centering\includegraphics[angle=270, width=\columnwidth]{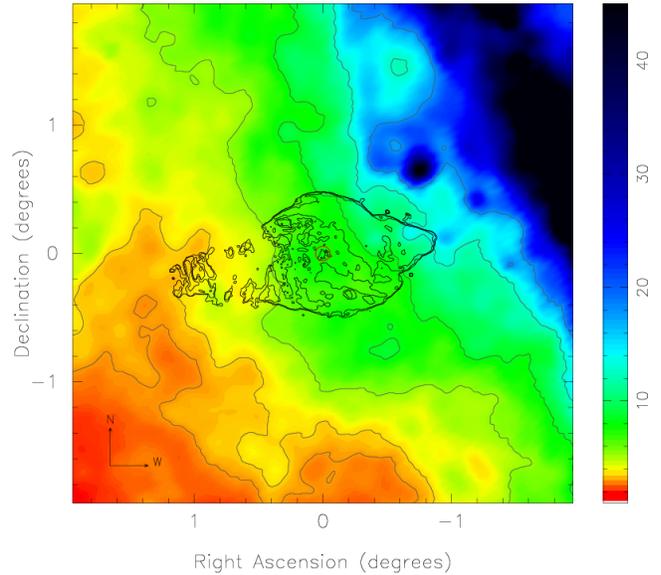}
  \caption{Colour-scale image shows the Galactic extinction ($A_V$)
    towards the W50/SS\,433 system from the IRAS/COBE all-sky survey
    \citep{sch98}. Colour scale corresponds to visual extinction in
    magnitudes and it is centred on SS\,433's position (given by the
    red circle). Grey contours on the colour image are 3, 3.6, 5.4,
    8.4, 12.6 and 18~mag. The radio continuum emission at 1465~MHz is
    shown as black contours \citep[data from ][]{dub98}. Radio
    contours are 12, 14 and 20~mJy~beam$^{-1}$. }
  \label{fig:dust}
\end{figure}

Fig.~\ref{fig:dust} shows the map of Galactic extinction in the
direction of the SS\,433/W50 complex. We constructed this extinction
image by converting the colour excess map of \citet{sch98} into an
estimate of the reddening, assuming a selective extinction, $R_V
\equiv A_V/E(B-V)$, equal to the average value for the Galaxy
\citep[i.e., $R_V=3.1$,][]{car89}, where $E(B-V)$ is the colour
excess. It is clear from the image that the $\sim$7.8~mag of
extinction is consistent with the interstellar absorption gradient.
The extinction estimate from this map towards SS\,433's position is
$A_V = 7.813$~mag. Although the value for $E(B-V)$ is quite accurate,
the choice of $R_V$ is not. The intrinsic error introduced by $R_V$
implies that the accuracy of the extinction estimate is not better
than to a tenth of a magnitude. Therefore, we corrected our spectra
(whose reduction is described in Section~\ref{sec:obs}) for $A_V =
7.8$~mag of interstellar absorption using the reddening law of
\citet{car89}, before calculating the Balmer decrements.

In this paper we decompose the profiles of the stationary emission
lines, H$\alpha$ and H$\beta$, with a number of Gaussian components
(Section~\ref{sec:decomposition}).  Each model component is identified
with its corresponding emitting origin. In Section~\ref{sec:results}
we analyse the physical conditions of each component present in
SS\,433 and its behaviour as a function of orbital and precessional
phases, via the Balmer decrement. In Section~\ref{sec:size} we present
our calculations for the accretion disc radii and compare our findings
with discs from other objects that share some common features with
SS\,433.

\section{Observations and data reduction}\label{sec:obs}

We present optical spectroscopy data covering most of a precessional
cycle of SS\,433's jet-axis (and presumably therefore accretion
disc). The data were acquired with the Supernova Integral Field
Spectrograph (SNIFS) at the University of Hawaii 2.2-m telescope by
the Nearby Supernova Factory \citep{snifs}. The spectrograph is
composed of two modules, one for blue wavelengths covering the region
between 320~nm to 560~nm with a resolution of $\sim$1000 at 430~nm,
and a module for red wavelengths (520~nm to 1000~nm) with a resolution
of $\sim$1300 at 760~nm. The data consist of 38 spectra (considering
blue and red as one spectrum). The data set can be subdivided in two
groups, one spanning from 2006 April 27 to July 30 (precessional
phases $\psi_{\rm pre} \in [0.7,1.3]$), and another one between 2006
October 1 and November 10 (precessional phases $\psi_{\rm pre} \in
[0.67,0.95]$). We use the convention in which orbital phase
($\phi_{\rm orb}$) zero is when the donor star is eclipsing out the
compact object \citep{gor98}. Precessional phases are calculated based
on the ephemeris reported in \citet{fab04}, where precession phase
zero is when the moving jet lines are maximally separated hence the
inclination of the jet axis with our line-of-sight attains a minimum,
i.e., it corresponds to maximum exposure of the accretion disc to the
observer.

The data were reduced by the Nearby Supernova Factory group through
their standard pipeline. Frequent observations of standard stars and
the arc lamp were performed during each night to perform flux
calibration and provide an accurate estimate of the wavelength
axis. Wavelength calibration was carried out by fitting a high-order
polynomial to the lamp spectra. Standard star division was applied to
each frame in order to flux calibrate the data. Telluric features were
not removed. All the subsequent analyses were carried out using the
Perl Data Language \citep[\texttt{http://pdl.perl.org,}][]{gla97}.

\begin{figure}
%   \centering\includegraphics[viewport=30 51 393 535, width=.9\columnwidth]{figs_pdf/paper_fit_example.pdf}
  \centering\includegraphics[width=.95\columnwidth]{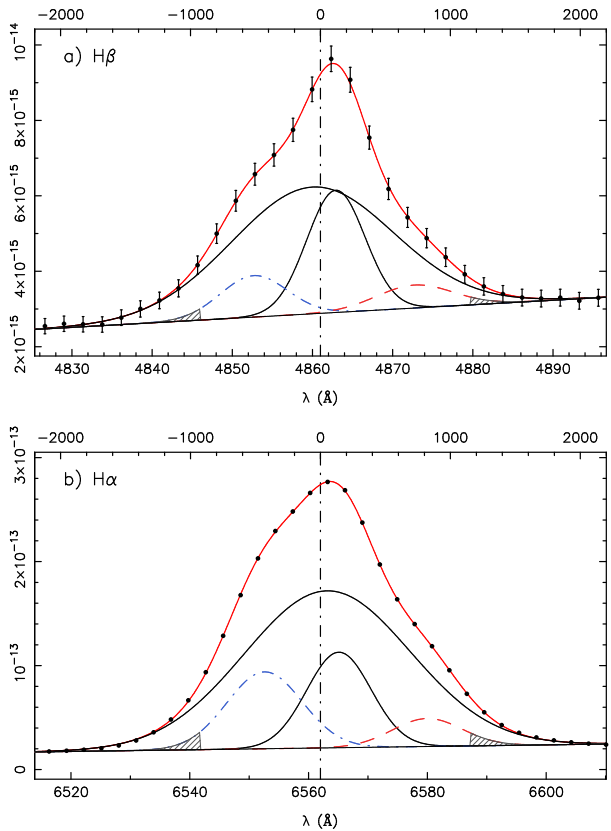}
  \caption{Examples of H$\beta$ (upper panel) and H$\alpha$ (lower
    panel) so-called stationary emission lines. Both spectra were
    acquired on MJD~53009.23, which corresponds to $\psi_\mathrm{pre}
    = 0.67$ and $\phi_\mathrm{orb} = 0.69$ (for a description of the
    convention used please see text in Section~\ref{sec:obs}). The
    different components are explained in the text. The narrow blue-
    (dot-dashed line) and red-shifted (dashed line) Gaussian
    components correspond to the accretion disc's flow. The top
    $x$-axis corresponds to rest velocity in units of \kms\ and both
    $y$-axes are in flux density units of W~m$^{-2}$~\mum$^{-1}$. The
    broad wind component has a FWHM of $\sim$1460~\kms\ and
    $\sim$1500~\kms\ for H$\beta$ and H$\alpha$, respectively. The
    error bars correspond to 3$\sigma$, which are about half the
    symbol size in the case of the H$\alpha$ line. The shaded area
    represents that which we consider to be the wings of the accretion
    disc profile (see Section~\ref{sec:size}).}
  \label{fig:examples}
\end{figure}

\subsection{Fitting procedure}

Examples of the reduced spectra in the vicinity of the H$\alpha$ and
H$\beta$ lines are shown in Fig.~\ref{fig:examples}. This figure also
shows the best combination of Gaussian components fitted to those
spectra. The blue spectra were rebinned to the spectral resolution of
the red spectra (130~\kms~pixel$^{-1}$). The underlying continuum was
subtracted by fitting a low-order polynomial (linear in most cases and
only in some cases parabolic). The best-fitting parameters were
determined by $\chi^2$ minimisation in the pixel space. We explain the
fitting model and the deconstruction of these profiles in the next
section.

\section{Decomposition of stationary lines}\label{sec:decomposition}

The stationary emission lines in SS\,433 possess complex and variable
profiles \citep{kmb08,seba09}. \citet{fal87} fitted the stationary
H$\alpha$ line profile of their low-resolution spectra using three to
four Gaussian components. Their best fit consisted of one broad
component plus three narrower components. \citet{kmb08} carried out
decomposition of the H$\alpha$ line using higher-resolution data. They
found one broad component (FWHM~$\sim 700$~\kms) whose width was
observed to decrease with precessional phase (i.e., as the jets become
more in the plane of the sky) identified as the accretion disc wind,
and two narrower, red- and blue-shifted, components which were
stationary in wavelength, being radiated from a glowing circumbinary
ring. In the data presented in that paper, there was no evidence of
the accretion disc lines being apparent, in contrast with their
persistent appearance in the data presented here.

Recently, \citet{seba09} were able to model the complex
Brackett-$\gamma$ (Br$\gamma$) stationary line with usually 5 Gaussian
components: a broad wind component (FWHM~$\sim\!1500$~\kms), two
Gaussian components accounting for the circumbinary ring and they also
found the presence of two components being radiated by matter
spiralling in the accretion disc. The presence of SS\,433's accretion
disc in the near-infrared line seems to be persistent and their
separation implies a circular velocity between 500 and 700~\kms. We
modelled our optical data following a similar approach.

Following the analysis done by \citet{seba09}, the H$\alpha$ and
H$\beta$ stationary lines were fitted with four Gaussian components in
order to account for the complexity of the profiles. Although two
narrow components have been used in the past to represent the split
lines from the circumbinary ring \citep{kmb08, seba09}, because of the
low resolution of the spectra we present in this paper
($\sim$130~\kms) just one central narrow component was needed to
represent the contribution from the
circumbinary. Fig.~\ref{fig:examples} shows examples of fitted
profiles.

\begin{figure*}
  \includegraphics[width=.61\textwidth]{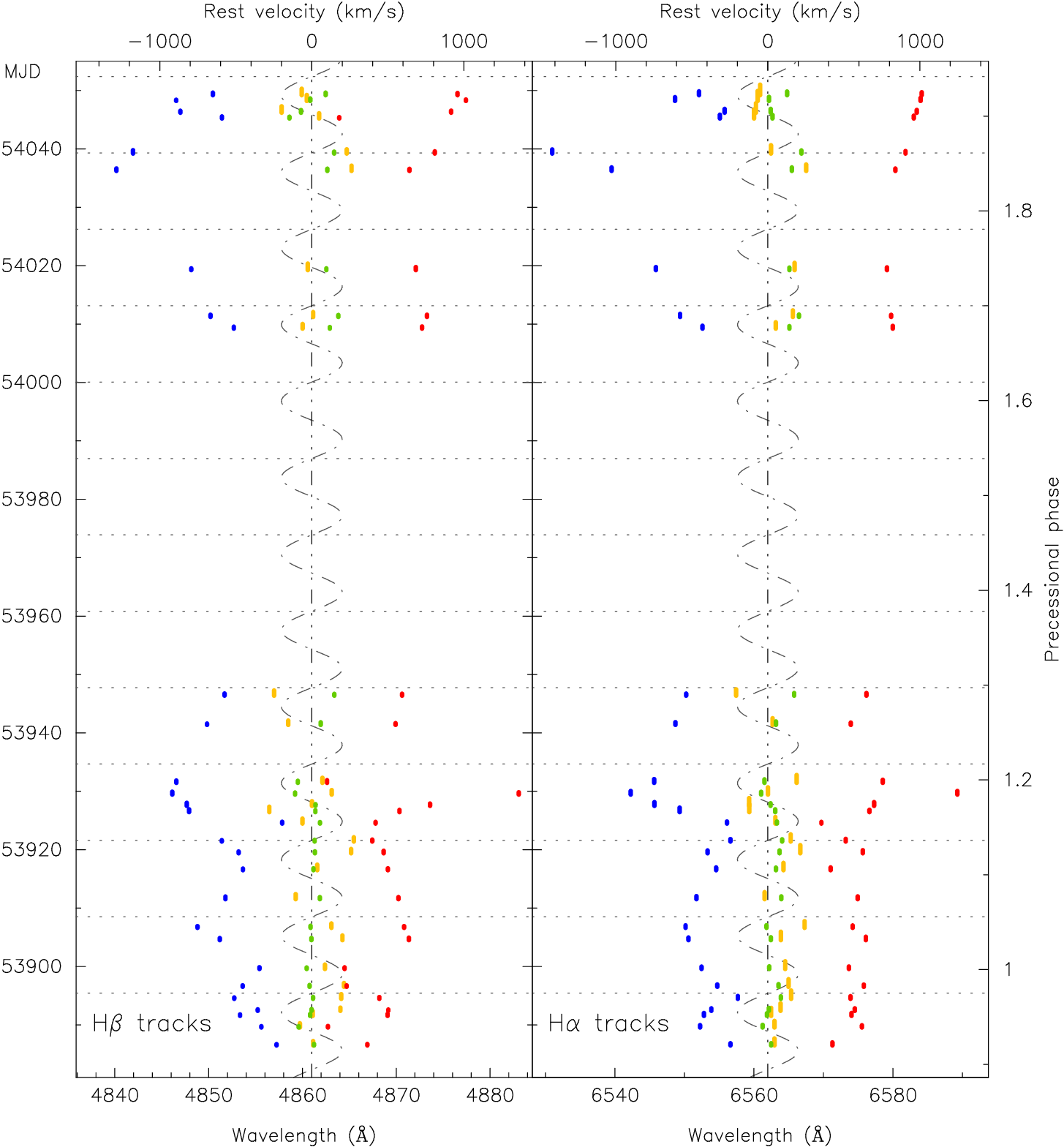}
  \hfill
  \includegraphics[width=.35\textwidth]{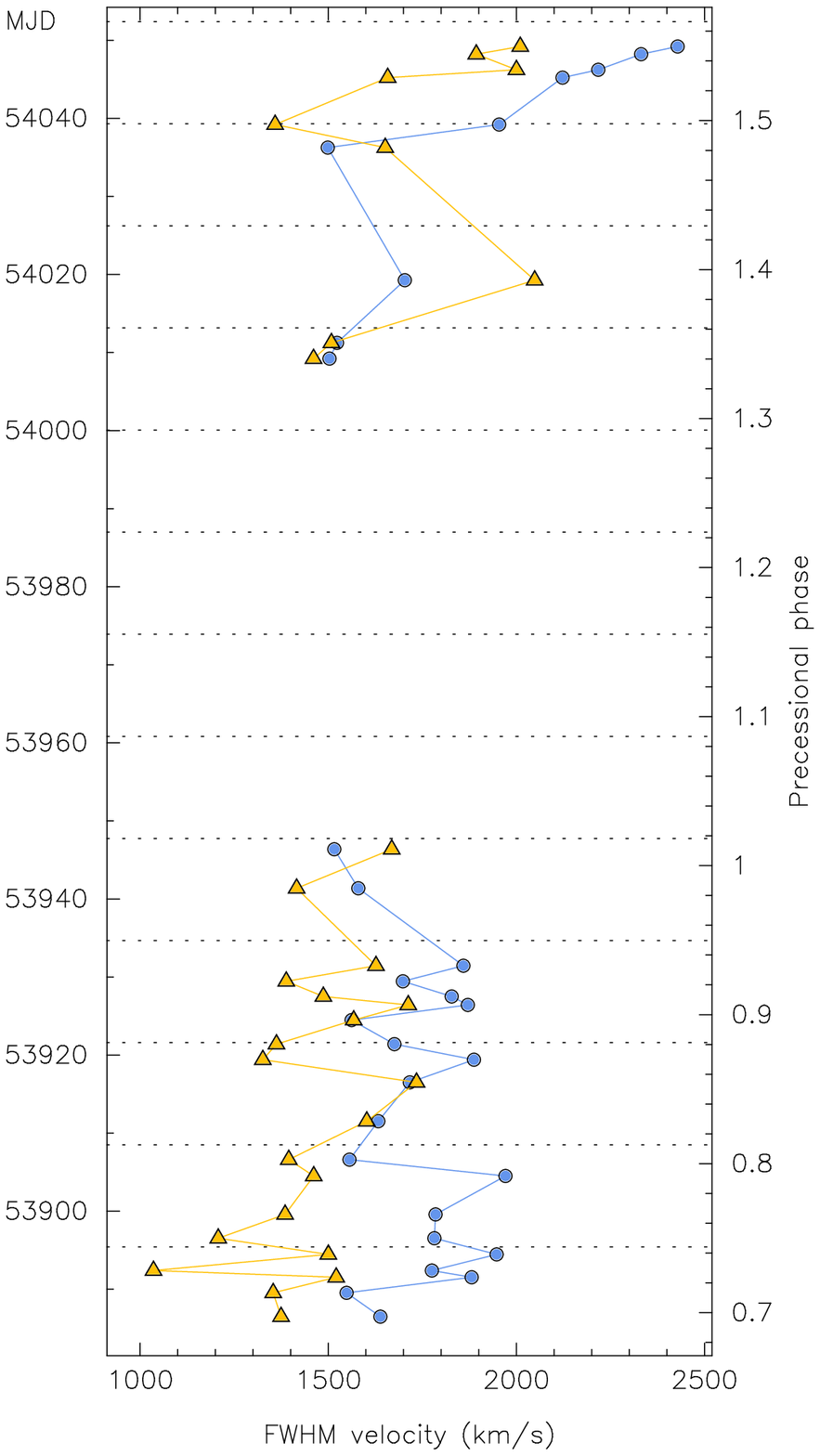}
  \caption{Tracks of the fitted centroids for each component of the
    stationary H$\beta$ ({\it left}) and H$\alpha$ ({\it middle})
    lines as a function of modified Julian day ($y$-axis, increases
    vertically). The tick mark heights are proportional to the FWHM of
    each component. Red and blue ticks correspond to the accretion
    disc lines, the larger black ticks represent the disc wind while
    the light blue ones represent the circumbinary ring. The top
    $x$-axis corresponds to rest velocity in units of \kms\ for an
    assumed systemic velocity of zero. The dot-dashed sinusoidal
    depicts the orbital motion of the binary. The dotted horizontal
    lines correspond to epochs when the orbital phase is zero. The
    plot on the {\it right} hand side corresponds to the evolution of
    the FWHM of the wind component of the stationary H$\beta$ (gold
    triangles) and H$\alpha$ (blue circles) lines.}
  \label{fig:tracks}
\end{figure*}

The presence of a P Cygni feature in the stationary lines at certain
precessional phases has been noted by several authors
\citep{cra81,fili88,gies02}. An absorption feature in the blue wing of
the stationary line profile would indeed complicate the analysis and
it would have to be taken into account using models of outflowing
winds \citep[e.g.,][]{cas79}. We detected the P Cygni absorption
signature at the epochs $\psi_{\rm pre} \in [0.7,0.85]$ and we have
excluded these data from our analysis.

The reason why we identify the accretion disc with two separate
components representing each peak of the disc's profile is based on
the model of emission line formation in accretion discs presented by
\citet{hor86}, in which the velocity of the outer regions of the disc
is represented by half the velocity separation of the line peaks
\citep{hor86}. This is caused by the Doppler effect and the rotational
motion of the matter within the disc. Therefore, the speed with which
the radiating material spirals within the accretion disc corresponds
to half the difference between the speed at which those lines are
moving (given by the relative position of the peaks), under the
assumption that the fitted centroids correspond to the tangent
speed. This reveals material that is orbiting in the potential well at
speeds of about $\gtapprox$600~\kms\ (see blue and red tracks in
Fig.~\ref{fig:tracks}).

Fig.~\ref{fig:tracks} does not reveal a clear signature of the
expected motion for the accretion disc centroid and certainly nothing
corresponding to the \citet{fab90} sinusoidal plot that assumes a
circular orbit.  It is not possible to identify what would be the
correct phasing of the motion of the disc due to lack of knowledge
about the circularity or eccentricity of the orbit. For certain
eccentricities, there are not necessarily two well-defined peaks, for
example.  A detailed analysis of this problem requires much more
long-term monitoring and sophisticated modelling.

Fig.~\ref{fig:tracks} (left and centre panels) clearly shows that the
accretion disc lines appear slower at orbital phase zero. This is when
the donor star obscures the inner (faster) region of the accretion
disc. This is seen most clearly in H$\alpha$. Additionally, there
appears to be an episode around MJD~53930, which corresponds to
orbital phase 0.5, where the accretion disc lines seem to be rotating
faster than usual and may simply correspond to a time when we have an
unimpeded view into the inner, hence faster, part of the accretion
disc. The right panel in Fig.~\ref{fig:tracks} shows the evolution of
the FWHM of the wind components of the stationary H$\alpha$ and
H$\beta$ lines. This panel also shows that the wind lines appear to
get broader at longer wavelengths, which is consistent with previous
near-infrared observations of the Br$\gamma$ line, where the wind
component has been reported to be broader than in H$\alpha$
\citep{seba09}. It is remarkable that near the end of the data set we
present in this paper, the width of the accretion disc wind attains a
velocity of almost 2500~\kms, which translates into a factor 1.7
increase in the 10$^{-4}$~\Msun~yr$^{-1}$ mass-loss rate estimated
from recombination line fluxes by \citet{seba09}.

\section{Results}\label{sec:results}

\subsection{Basic picture}

Modelling of the stationary Balmer emission lines, H$\alpha$
\citep[this work and ][]{kmb08}, H$\beta$ (this work) and Br$\gamma$
\citep{seba09}, yields a clear and consistent interpretation of the
radiating components present in the SS\,433 system. The main
components of the system, as inferred from the decomposition of the
stationary line complexes, are: a rotating accretion disc, a fast
outflowing wind that accounts for most of the mass loss in the system
and a circumbinary ring probably being fed from the L2 point. The
moving emission lines from the jets sometimes blend with the
stationary ones at around $\psi_{\rm pre} \sim 0.3$ or 0.4,
contributing significantly to the intensity variation and profile
shape. Moreover, care must be taken since the intensities and profiles
of the stationary lines vary strongly during flares \citep[Blundell et
al. in prep., ][]{mar84}.

\subsection{Balmer decrement of stationary lines}

\begin{figure}
%   \centering\includegraphics[viewport=44 155 540 612, angle=270,width=\columnwidth]{figs_pdf/paper_triple_balmer_decrements.pdf}
  \centering\includegraphics[angle=270,width=\columnwidth]{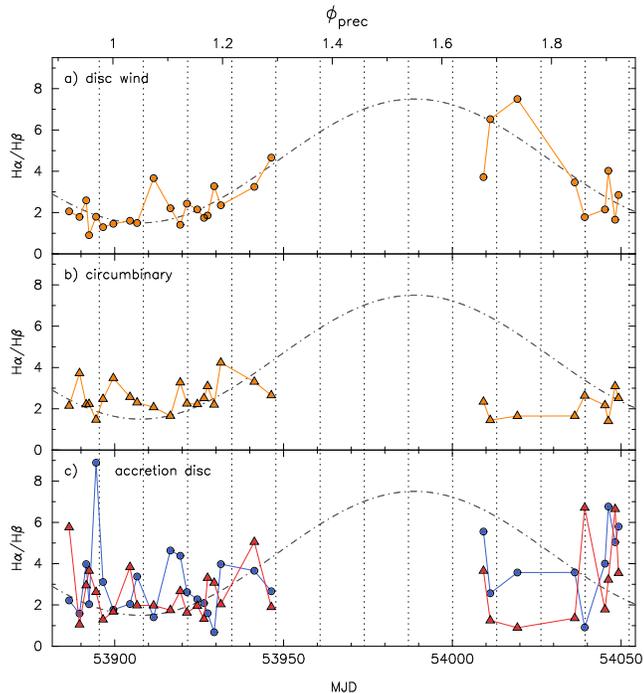}
  \caption{The Balmer decrement, H$\alpha$/H$\beta$, as a function of
    precessional phase (top $x$-axis) or modified Julian day (bottom
    $x$-axis), for the accretion disc wind (top), circumbinary ring
    (middle panel) and accretion disc lines (bottom). The dot-dashed
    line corresponds to the best fitted sinusoidal to the disc wind
    decrements, with a fixed period of 162~days. In the bottom plot
    the red- and blue-shifted accretion disc lines are represented by
    solid triangles and circles, respectively. The vertical dotted
    lines represent times at which the orbital phase of the binary is
    zero.}
  \label{fig:dec}
\end{figure}

The hydrogen Balmer lines show very variable Balmer decrements in
SS\,433. Fig.~\ref{fig:dec} shows that the computed Balmer decrements
for the stationary emission-lines scatter from $1
\ltapprox$~H$\alpha/$H$\beta$~$\ltapprox 8$. In the interpretation of
these results we use the calculations carried out by \citet{dra80} who
modelled the emission-line spectrum from a slab of hydrogen at high
densities, by taking into account important mechanisms such as
collisional excitation and de-excitation, as well as self-absorption
processes. It is important to chose reasonable temperature and optical
depth ($\tau_{\rm L\alpha}$) parameter values in order to apply their
calculations to SS\,433. We mainly utilise the decrements computed for
a model with $T_{\rm e} \gtapprox 10^{4}$~K. For a
theoretically-modelled accretion flow, Kramer's approximation for the
optical depth yields $\tau_{\rm L\alpha}=4\times 10^{4}$
\citep{dra80}. Since decrements for this exact value are not reported
in their paper, we use $\tau_{\rm L\alpha}\sim 10^{5}$
instead. According to \citet{dra80} these ratios ($1
\ltapprox$~H$\alpha/$H$\beta$~$\ltapprox 8$) then imply up to four
orders of magnitude variation in density throughout the precession
period. This large inferred variation is not a problem for plausible
physical models.  We will analyse and discuss the Balmer decrements
calculated for each component of SS\,433 in the following
sub-sections.

\subsubsection{Circumbinary disc decrements}

The Balmer line ratios of the components corresponding to the
circumbinary excretion disc observed by \citet{kmb08} are shown in the
middle panel of Fig.~\ref{fig:dec}. They are roughly constant
throughout precessional phase with an average value of
H$\alpha$/H$\beta= 2.45 \pm 0.69$ (1~$\sigma$ scatter). This implies a
moderate electron density of $\log N_{\rm e}\simeq 11.5$,
approximately constant throughout the precessional cycle and given the
similarity of this Balmer decrement to the Case B canonical value of
$\sim$2.8, this is consistent with no attenuation from the accretion
disc wind, in accordance with the {\em circumbinary} nature of this
emitting region \citep{doo09}.

\subsubsection{Accretion disc wind decrements}

Fig.~\ref{fig:dec} shows that in the case of the accretion disc wind,
the Balmer decrements have a very clear precessional phase dependence,
showing a clear tendency towards lower electron density (steeper
decrements) as the jet axis becomes more in the plane of the sky. This
dependence of density on direction is evidence for the polloidal
nature of the wind. The decrement values for the disc wind vary
between quite flat H$\alpha$/H$\beta = 1.43 \pm 0.28$ between
precessional phases 0.95--0.05 and steep H$\alpha$/H$\beta = 5.30 \pm
1.75$ for phases 0.65--0.84. These values correspond to variations in
electron density of $\log N_{\rm e}\simeq 13$ and $\log N_{\rm
  e}\simeq 10$, respectively.

\subsubsection{Accretion disc decrements}

The accretion disc lines are the ones that show the most dramatic and
scattered fluctuations in decrement. Both lines, H$\alpha$ and
H$\beta$, show a similar level of scatter. The average decrement for
the red and blue components are H$\alpha$/H$\beta = 2.78 \pm 1.56$ and
H$\alpha$/H$\beta = 3.34 \pm 1.80$, respectively. Inverted Balmer
decrements can be seen at some epochs, implying very high densities
and possibly an enhanced inflow of matter from the companion onto the
accretion disc. The decrement of the accretion disc itself appears to
be dominated by fluctuations in the instantaneous wind density rather
than being dominated by any precessional phase dependence.

\section{Size of the accretion disc}\label{sec:size}

A direct consequence of \citeauthor{hor86}'s \citeyear{hor86} model of
the emission line profile formed in an accretion disc is that the
emission close to the inner radius of the disc contributes to the
outer wings of the double-peaked profile, while the regions nearer the
outer radius provide the emission for the peaks \citep[see fig.~1
in][]{hor86}. By considering this along with the assumption that the
velocity of the matter spiralling in the accretion disc is Keplerian,
we can estimate the size of the disc \citep{mas00}.

For a test particle in Keplerian motion at a distance $r$ from a
central object of mass $M_{\rm bh}$, the Keplerian speed projected
onto the plane of the sky is
  
\begin{equation}
  V_{\rm Kep} = \sqrt{\frac{G M_{\rm bh}}{r}} \, \sin i,
  \label{eq:vel}
\end{equation}

\noindent where $i$ is the angle between the normal to the disc and
our line-of-sight. The disc is edge-on when this inclination angle is
$90^{\circ}$ and pole-on when $i=0^{\circ}$. In our decomposition of
the Balmer lines' profiles, the Keplerian velocity of the gas at the
outer radius $R_{\rm out}$ of the disc is given by half the separation
between the peaks of the two Gaussian components representing the
accretion disc, while the Keplerian velocity of the gas closer to the
disc inner radius $R_{\rm in}$ corresponds to half the separation
between the red wing of the red-shifted component and the blue wing of
the blue-shifted component (see hatched area in
Fig.~\ref{fig:examples}).

The most suitable epoch to measure the inner and outer radii of the
accretion disc is when the disc is in between the companion star and
the observer (orbital phase 0.5) and when the disc is not totally
edge-on (i.e., $\sin i \sim 1$) at precessional phase $\psi_{\rm
  prec}$ close to 0.33 or 0.66. The closest spectrum in our dataset to
those epochs was taken on MJD~54009 which corresponds to $\psi_{\rm
  prec} = 0.67$ and $\phi_{\rm orb}=0.69$. This spectrum is depicted
in Fig.~\ref{fig:examples}.  Both the half wing separation and the
half peak separation for the accretion disc lines were measured from
this spectrum.

Half the separation between the peaks of the fitted Gaussian
components of the disc yields a Keplerian velocity for the outer part
of the disc of $\sim$620~\kms, measured from both the H$\alpha$ and
H$\beta$ emission. The Keplerian velocity of the inner part of the
disc was determined by measuring the positions in velocity of the line
wings (centre of the hatched areas in Fig.~\ref{fig:examples}) by
inspection. Both H$\alpha$ and H$\beta$ present wing velocities of
$\sim$1150~\kms.

The central compact object in SS\,433 is most likely to be a stellar
mass black-hole, rather than a neutron star
\citep{gies02,fab04,lop06,kmb08}.  It is evident from
Equation~\ref{eq:vel} that large black-hole masses imply larger outer
radii for the accretion disc. The total mass of the accretion disc
plus the compact object has been calculated by \citet{kmb08} and it
attains 16~\Msun. However, we point out that there has been a large
dispersion reported for the component masses of SS\,433
\citep[e.g.,][]{lop06}. We will assume that the mass of the black hole
itself is of the order of 10~\Msun.

\begin{figure}
  \centering\includegraphics[angle=270,width=\columnwidth]{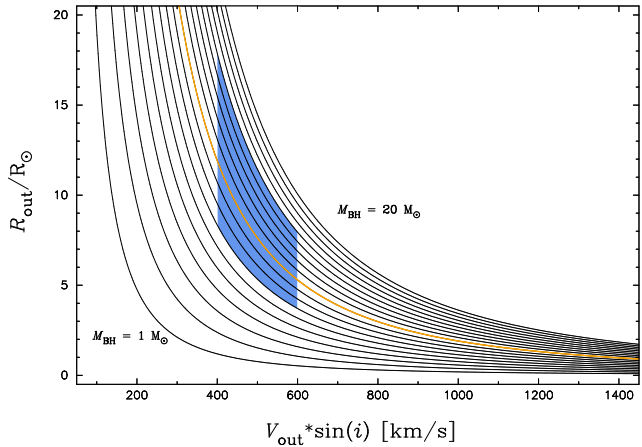}
  \caption{Variation of the outer radius of the accretion disc as a
    function of the Keplerian velocity of the outer edge of the disc
    (half the separation of the line peaks, in \kms), for different
    values of the mass of the compact object. The solid area
    represents the range of parameters that we think are relevant to
    SS\,433.  The thick orange line corresponds to \mbox{$M_{\rm bh} =
      10~\Msun$}.}
  \label{fig:rout}
\end{figure}

The Keplerian velocity for the inner region of the disc of 1200~\kms\
implies that the accretion disc's inner radius is \mbox{$R_{\rm in}
  \sim 1.4~M_{10}$~\Rsun}, where $M_{10} = M_{\rm bh}/10\,\Msun$.  The
estimated value for $R_{\rm in}$ corresponds to an upper limit to the
inner radius of the accretion disc at the time of observation. This is
a direct consequence of the high obscuration presented in the system
due to the accretion disc wind, which hinders the emission propagating
from the inner parts of the disc. Additionally, the determination of
the full-width at zero intensity (FWZI), or distance between the
wings, could be imprecise sometimes and it could add certain degree of
uncertainty to the $R_{\rm in}$ value.  Also, it is important to note
that the inner radius of an accretion disc is known to recede as a
consequence of changes in the accretion rate onto the compact object,
for a range of masses of the compact object.

On the other hand, the distance between the peaks of the accretion
disc components depicted in Fig.~\ref{fig:examples} yields a velocity
for the disc outer region of about 620~\kms, which if Keplerian yields
$R_{\rm out }\sim 5~M_{10}$~\Rsun. Fig.~\ref{fig:rout} shows how
$R_{\rm out }$ varies as a function of the Keplerian velocity of the
outer edge. This is fully consistent with a previous estimate of the
overall size of the system, whose semi-major axis of the orbit has
been reported to be $\sim$80~\Rsun\ \citep{seba09}.

The Keplerian velocity of the outer region of the accretion disc has
been reported to be lower in the near-infrared Br$\gamma$ line
\citep{seba09} than in the optical lines analysed in this paper. This
is consistent with observations of other accretion discs such as the
one in WZ Sge. By analysing Doppler tomography maps, \citet{ski00}
found that the Br$\gamma$ emission is being emitted from a region
further away from the central object, implying that this emission is
more representative of the outer regions of the disc. This is probably
the case in SS\,433 since the peak separation in Br$\gamma$ is smaller
than in the Balmer lines. The reported half separation of the peaks of
$\sim$500~\kms\ by \citet{seba09} implies an outer radius $R_{\rm out}
= 8~M_{10}$~\Rsun.

The black-hole mass does not affect the inner to outer radius ratio,
since $R_{\rm in}/R_{\rm out} = \left(V_{\rm out}/V_{\rm
    in}\right)^2$.  In the case of SS\,433's disc, $R_{\rm in}/R_{\rm
  out} \approx 0.2$ (for an outer radius $R_{\rm out} \sim
8~M_{10}$~\Rsun).

The size of SS\,433's accretion disc is quite large compared with
other close binary systems. \citet{mas00} determined the size of
WZ~Sge to be at most 0.43~\Rsun\ for a white dwarf with the
Chandrasekhar mass. This is about 20 times smaller than SS\,433's
disc. The $R_{\rm in}$:$R_{\rm out}$ ratio is approximately the same
in both these systems.

\section{Conclusions}

We have presented optical spectroscopy of the microquasar SS\,433
covering a significant fraction of the precessional cycle of its
accretion disc. The components of the prominent stationary H$\alpha$
and H$\beta$ lines have been identified as arising from three emitting
regions: an accretion disc wind which is super-Eddington
\citep{seba09}, in the form of a broad component accounting for most
of the mass loss in the system, a circumbinary disc of material
probably excreted through the binary's L2 point, and the accretion
disc itself as two persistent components, having an outer region
Keplerian velocity of $\gtapprox 600$~\kms.

A direct result of this decomposition using our UKIRT data published
in \citet{seba09} was the determination of the accretion disc size,
whose outer radius attains 8~\Rsun, for an assumed black hole mass of
10~\Msun. With the data presented in this paper we determined the
accretion disc inner to outer radius ratio in SS\,433, $R_{\rm
  in}/R_{\rm out}$ to be $\sim$0.2, independent of the mass of the
compact object.

The Balmer decrements, H$\alpha/$H$\beta$, were extracted from the
stationary emission lines for each component of the system. The
decrement of the circumbinary ring seems to be quite constant
throughout precessional phase, implying a fairly constant electron
density of $\log N_{\rm e}\simeq 11.5$ for the circumbinary disc. The
accretion disc wind shows larger changes in its decrements as a
function of precessional phase, implying variations in its density,
$N_{\rm e}$, between $10^{10}$ and $10^{13}$~cm$^{-3}$.  Thus, the
physical parameters of the gaseous components imply rather dense
environments emitting the Balmer lines.

\section*{Acknowledgments}

We thank the referee for a careful reading of the paper. We are very
grateful to the Sciences and Technology Facilities Council ({\sc
  stfc}) and Conicyt for the award of a {\sc STFC}-Gemini
studentship. We are specially grateful to Greg Aldering, the Nearby
Supernova Factory collaboration and the University of Hawaii for their
generosity in allowing us to obtain some spectra on SS\,433 during
interstitial time within their main supernovae follow up
campaign. K. B. thanks the Royal Society for a University Research
Fellowship.

\bsp

\label{lastpage}

\end{document}